# Form, function, mind: what doesn't compute (and what might)

Stuart A. Newman*
*New York Medical College, Valhalla, New York 10595 USA*

___

ABSTRACT

The applicability of computational and dynamical systems models to organisms is scrutinized, using examples from developmental biology and cognition. Developmental morphogenesis is dependent on the inherent material properties of developing tissues, a non-computational modality, but cell differentiation, which utilizes chromatin-based revisable memory banks and program-like function-calling, via the developmental gene co-expression system unique to metazoans, has a quasi-computational basis. Multi-attractor dynamical models are argued to be misapplied to global properties of development, and it is suggested that along with computationalism, dynamicism is similarly unsuitable to accounting for cognitive phenomena. Proposals are made for treating brains and other nervous tissues as novel forms of excitable matter with inherent properties which enable the intensification of cell-based basal cognition capabilities present throughout the tree of life.

# 1. Introduction

Organisms develop, thrive, behave, reproduce, and evolve. Computers compute. Does it make any sense to conflate these activities? Organisms were first described as machines during a time (the 18$^{th}$ century) when industrialization was making machines ubiquitous, and the principles of mechanics were the scientific cutting edge [1]. In the mid-20$^{th}$ century, the invention of computers inspired a new conception of organisms and their vital processes (e.g., Schrödinger's chromosomes as "code law and executive power…in one" [2]). Also around then, animal brains ceased being likened to 1920s-style telephone switchboards [3] and instead became identified with the central processing units of digital computers whose rules and routines – i.e., programs – were hypothesized to be arrays of connections in the respective species' nervous systems. The information encoded in this software was presumed to generate behavior, perception, and cognition when executed by the hardware of the brain's tissue [4, 5].

Development was thought of similarly: the 1970s and 80s were the heyday of the notion of species-specific "genetic programs" directing the embryo's cells through a sequence of gene expression states to generate the spatiotemporally organized body of a viable individual [6]. While the computer concept of the brain has been updated with the advent of connectionist processors (themselves based on newer ideas of how nervous systems work: "neural networks" with no predetermined programs) [7], developmental biology has remained largely dominated by the genetic program idea. The main exception is cell differentiation, which has been modeled by multistable dynamical systems [8], but these models have been questioned [9].

As with the artist Albrecht Dürer's famous rendering of a rhinoceros as an exotic beast clad in a suit of armor (Fig. 1), it is natural to try to explain unfamiliar things using concepts and materials at hand (see Gombrich [10]). The hallmark of authentic scientific thought, however, is not the assimilation of the unknown to the known, but the less reductive program of situating observations in contexts and frameworks in which we have confidence and inventing new, compatible concepts where necessary.

An example is thermodynamics, a set of principles derived from the study of steam engines and later, chemical reactions. This theory, Einstein asserted, "within the framework of applicability of its basic concepts will never be overthrown" [11]. Further, it is as broadly applicable as any in physics (only quantum mechanics claims equivalent universality).

Newtonian mechanics applies to a narrower domain, but its consistency with thermodynamics is demonstrated by the derivation of observables like entropy and heat capacity (which have no meaning in terms of Newton's laws of motion) by probabilistic analysis of ensembles of systems of particles subject to those laws [12]. This is not reduction of thermodynamics to Newtonian physics, but a confirmation that these two partial characterizations of matter exist in the same reality.

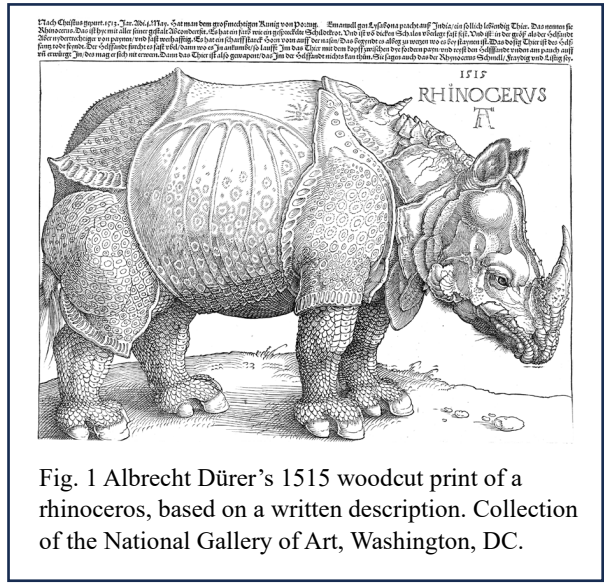

Fig. 1 Albrecht Dürer's 1515 woodcut print of a rhinoceros, based on a written description. Collection of the National Gallery of Art, Washington, DC.

More generally, the natural world is characterized by things composed of qualitatively distinct forms of matter. While a materialist view requires that different kinds of matter are consistent or even mutually transformable at the boundaries or shared domains of their existence, this does not mean that their characteristic properties are reducible in any hierarchical sense. The reflectivity and fluidity of mercury, a unique form of matter, depend on and are reconcilable with, but are not the same as, the properties of mercury atoms, and the latter's properties are different in kind from the smaller particles that compose them. The distinctive characteristics of a "unitary" form of matter (defined as a material having characteristic *inherencies*, e.g., the propensity of water to form waves or vortices, and intensive properties, like viscosity) are restricted to a narrow range of spatial and temporal scales.

The foregoing introduces the perspective of this paper on the bases of biological form and function and their relation to computation. My main claim is that key aspects of living organisms, most prominently their morphogenesis (generation of form), and possibly aspects of their sensory and cognitive capacities, are expressions of the inherencies of the materials of which they are composed. Other features, including the differentiation of specialized cells during development and possibly retrieval of stored patterns in the brain, utilize quasi-computational storage, data addressing and function calling processes intrinsic to the cells that constitute the organisms' tissues.

Every tissue that contributes to forming the bodies (including brains) of multicellular organisms is composed of living cells, eukaryotic (nucleated) ones in these examples. (Prokaryotic cells also form assemblages – biofilms, streams, and fruiting bodies – with similarly nonstandard physical properties.) These multicellular materials have species-characteristic, generally uniform (though often time-dependent) modes of cell-cell attachment and degrees of relative mobility. This justifies their being considered (at suitable scales) unitary forms of matter.

Some properties of these multicellular materials are "generic," that is, shared with nonliving counterparts such as deformable and nondeformable solids and viscoelastic liquids [13]. This permits them to be investigated with experimental tools and mathematical models used to study nonbiological matter. But living materials can exhibit generic inherencies for different reasons from their nonliving counterparts. For instance, a material can sustain chemical gradients if it contains a concentrated substance that can spread out from its source. In nonliving systems, transport is usually by passive diffusion; in tissues it can be facilitated by activities of cells along the route. The material distribution curves can nonetheless be identical.

Another example pertains to liquid-like materials. The relevant physics pertains to systems whose subunits are simultaneously cohesive and randomly mobile [14]. The subunits of molecular liquids are reversibly bound to one another by van der Waals forces and they undergo thermally driven Brownian motion. For liquid-like tissues, where the subunits are cells, the corresponding determinants are cell-surface attachment molecules and undirected, cytoskeletally driven motility [15]. In such cases, where outcomes are similar for similar physical reasons, the inherencies of the living materials have been termed "biogeneric" [16].

The tissues of plants (discussed elsewhere [17, 18]), are a unitary form of matter, but they are not biogeneric in the same way as animal liquid-like tissues are. While the distinction is not absolute, the tissues of plant embryos and organ primordia are composed of non-motile but individually extensible subunits globally connected by signal-transmitting plasmodesmata (long cell extensions). This kind of material does not have any counterpart in the nonliving world; its inherent, and therefore recurrent, morphological motifs (e.g., buds, leaf-like structures, branches, floral organs) require *sui generis* explanations. While generic biological materials are more straightforward to study, the inherencies of non-generic ones can be characterized by suitable application of physical laws and experimental and computational methods [19-21].

The proposal that developing plant and animal tissues are unitary forms of matter and that their morphogenetic potentialities (including some evolutionary trajectories) are expressions of the inherencies of these materials, does not imply that all biological features can be understood in these terms. After body and organ forms arose (e.g., in metazoans), they continued to evolve. Even if the established forms remained in place, the originating morphogenetic processes could be altered, supplanted, or replaced [22-26]. When forms which were originally generated by biogeneric or other material inherencies no longer develop under their influence, the resulting motifs, e.g., layers, segments, limb digits, while reflecting the original causes, may resist analysis by simple physical models [27]. In such cases, diachronic explanations, taking evolutionary transformations from the hypothesized conditions of origin into account, are required [28].

But developmental processes that have not evolved too far from their original condition can continue to be studied using standard physical techniques and their properties interpreted by known physical laws. This has led to major successes in development and neurobiology [15, 29, 30]. In addition, environmentally induced plasticity of developmental outcomes can indicate the continuing efficacy in present-day embryos of biogeneric physical determinants [31-33].

There are cases, however, where effects that occur in biological systems are unprecedented, unlike anything seen in the nonliving world [34]. These phenomena call for new investigative techniques and (in the words of the early 20[th] century embryologist E.E. Just) "a physics and chemistry in a new dimension, in a new dimension … superimposed upon the now known physics and chemistry" [35]. Some subcellular materials, "biomolecular condensates" [36], are not only non-generic but are truly exotic, with properties that currently defy explanation in terms of chemistry and physics [37, 38]. Such condensates are amalgams of intrinsically unstructured proteins and nucleic acids, with roles throughout the cytoplasm and nuclei of eukaryotic cells. These include the constitution of membraneless hubs of coordinated gene expression [39] and the formation of the brain's synaptic connections [40, 41]. Although little is understood about them at present, biomolecular condensates could be a unitary form of matter with inherent properties that underlie their unique functional capabilities. At more macroscopic levels, nervous tissues, including the brain, might be composed of one or more novel unitary forms of matter with inherencies that explain their capabilities. This possibility will be discussed in Section 5, below.

Cells themselves, the subunits of all complex living materials, are different in kind from multicellular matter in any of its varieties. While they are chemically continuous with the nonliving world at the elemental level, they contain components such as biomolecular condensates whose properties, as mentioned, are not close to being understood. Cells are also internally organized, compartmentalized, and compositionally and functionally heterogeneous, and therefore not characterizable as unitary forms of matter with inherent properties. Some cell behaviors resemble computations, and thus potentially contribute to computation-like, rather than material inherency-driven, properties of organisms. These features will be discussed in Section 2, below, before addressing the determinants of development and cognition.

The subject of this paper is the alternative applicability of material inherency vs. computation-like effects as explanatory modes in multicellular biology, including development and cognition. Although it has been broached philosophically [42], few have suggested that all physical processes are computations, and less so that every physical entity is a computer. My default assumption is that biological entities are like other natural entities in that they reflect the inherent properties of the materials that constitute them. This implies that many or most living features do not arise from computations and the organisms or embryos that embody them are not implementing computer-like programs. As we will see, however, the constituent cells complicate things by introducing quasi-computational modalities to the properties of biomaterials.

**2. Cells as non-unitary forms of matter**

All cells contain macromolecules – proteins, nucleic acids, lipid and polysaccharides – which have long evolutionary histories. The roles of these molecules depend on their being organized into circuits and assembled into complexes, all the functions of which have meanings within the cell that are distinct from their purely physical and chemical properties. Further, DNA and RNA have symbolic or coding relationships between them, and with proteins [43]. While not constituting programs in the computational sense, these nucleic acid-protein correspondences are "addressable" during a cell's functioning. This leads cells (whether prokaryotic or eukaryotic) to change their composition conditionally (via gene regulation) in response to alterations in their environments. In addition, eukaryotic cells contain within their nuclear DNA-protein complex ("chromatin") DNA-binding proteins (histones) that undergo reversible modifications ("marks") in the form of condition-responsive acetylation and methylation which can activate or inhibit gene expression. This constitutes a system of revisable epigenetic memories of the cell's lifetime experiences [44]. This "write-read-rewrite" system represents an advance in eukaryotic cells relative to prokaryotes in their capacity for agential behavior and what has been termed "basal cognition" [45].

In the cells of Metazoa (animals), an evolutionary derivative of the eukaryotes, chromatin structure is even more elaborate, incorporating the previously mentioned biomolecular condensates in phase-separated domains containing expressed or quiescent genes. The uniquely versatile metazoan gene expression apparatus resulting from the confluence of these information-processing features renders animal embryos loci of computation-like activity in the development of cells, tissues and organs (see Section 4, below).

As internally heterogeneous and structured entities, cells are clearly not a unitary form of matter in the sense that these are encountered outside of life (e.g., various gases, liquids, and solids, Bose-Einstein condensates, black holes), and sometimes within it (developing plant and animal tissues). Cells are more like complex multicomponent entities such as lichens, ecosystems, and cities. But all of these are derivative of cellular life and none is as internally integrated as cells themselves. Abstract theories of the living state, e.g., autopoiesis [46] and the "organizational approach" [47], although frequently applied to whole organisms, are based on postulated irreducible properties of the cell: operational closure and organizational closure, respectively. There have been few plausible hypotheses advanced thus far for the coming-into-existence of such organizational arrangements, but Terrence Deacon's "autogenesis" seems to be the most promising [48].

One uncontroversial feature of any cell concept is that all cells, irrespective of their identity as prokaryotic, eukaryotic or (within the latter), metazoan, embody life-sustaining *functionalities*. These are processes such as motility, secretion, adhesion, sensation, excitability, detoxification, replication (in addition to the essential ones of nutrient uptake, metabolism, and biosynthesis) [49]. In multicellular organisms like animals, some of these cell functionalities are recruited during embryogenesis to generate specialized tissues and organs in coordination with material inherency-based morphological development [50].

I will illustrate this in the next two sections, first by describing the current understanding of the morphological development of animals. I suggest (contrary to the notion of the "genetic program for development") that the characteristic morphologies and cell distribution patterns of the members of this group are largely inherent to the multicellular matter they are made of. In the subsequent section, I provide evidence that the emergence of specialized cell and organ functions concomitant with morphological development occurs by different means: the quasi-computational (e.g., function calling [51]) capabilities of metazoan embryonic cells. In the penultimate section I speculate on how the "inherencies of matter" view might complement or partly supersede more familiar computationist theories in understanding how brains function in mediating sentience and cognition, while leaving space for some of the quasi-computational modalities employed in cell differentiation. I will conclude with a brief discussion of the realms of applicability of these different kinds of explanation in biology.

3. Animal forms as material inherencies

While the organization and functionalities of cells are enigmatic and have resisted explanation from first principles (i.e., prebiotic physics and chemistry) this is not entirely true for the properties of complex multicellular organisms. First, as indicated above, cell associations constitute forms of matter with understandable and even predictable physical properties. Second, the development of organisms from such materials may harness or repurpose existing features of the component cells, which though not explicable in themselves, preexisted the multicellular novelties. Third, the geological and fossil records, read in relation to the genetic divergence among extant organisms, provide insights on temporal order and scenarios for emergence of post-unicellular forms of life [52-54].

In this section, I consider the material bases of metazoan morphogenesis and cell pattern formation. Clusters of predifferentiated animal cells (i.e., cells of developing embryos or the primordia of organs) have liquid-like properties. In contrast, clusters of cells from nonmetazoan holozoans, the phylogenetically closest extant relatives of the animals, do not have the same material nature. The reason is that the evolutionary transition to animals (around 800 million years ago; [55]) was accompanied by the acquisition of a transformational protein novelty. While extant unicellular holozoans have, and ancestral ones are inferred to have had, surface proteins of the *cadherin* family, the cadherins of these organisms pass through the cell membrane but do not connect to the contractile cytoskeleton. The novelty in animals was the addition to the cadherin protein of such a connector. Now, uniquely among the eukaryotes and indeed all cellular lineages, tissues existed the subunits of which were simultaneously cohesive and independently mobile, the hallmark of the liquid state (reviewed in [50]).

The inherencies of the liquid-tissue state paved the way for all subsequent morphological evolution of the animals. That is, although all forms in the space of possibilities ("morphospace" [56]) of this form of biological matter appeared at once, they were readily accessible with minimal genetic change. The evolution of the animal body plans can be placed in rough order by a physics-based transit through this space [57].

Liquid droplets are spherical by default, which is the shape of the earliest metazoan fossils and the starting point of most developing embryos. Liquid-like materials can also *phase separate* establishing immiscible subregions, if their subunits have sufficiently different cohesive properties. Differentially adhesive populations of cells can serve as biogeneric counterparts of oil and water molecules as they spontaneously arrange into embryonic tissue layers (reviewed in [15]).

Another property inherent to liquids with certain kinds of subunits is the formation of interior spaces. Subunits with nonuniform properties ("polar") spontaneously rearrange, leading to hollow or asymmetric forms. Cells in animal embryos have two main ways of becoming polar, each affecting the shape or form of the respective tissues (reviewed in [58]). *Apicobasal polarization* (ABP), in which the adhesive properties of cells become nonuniform around their perimeters, causes a tissue mass to generate a lumen or internal cavity, analogously (and by a similar physical mechanism) to micelles that form when amphipathic molecules are mixed with water. *Planar cell polarization* (PCP), in which the shapes of cells become nonisotropic, can lead to their mutual intercalation, causing the tissue mass to elongate. Similar reshaping occurs (for similar reasons) in a droplet of a molecular liquid crystal [59].

There is no PCP in sponges and placozoans, the animals with the simplest body plans. It is unclear whether these groups branched off the animal phylogenetic tree before the relevant pathways evolved, or the pathways were lost in those lineages, foreclosing the associated tissue changes (reviewed in [50]). All other animal groups (referred to as "eumetazoans") utilize PCP. These organisms can also form a *basal lamina*, a flexible planar sheet of solid extracellular material [60]. This structure, a departure from the liquid state of the tissues that generate it, permits the latter to adhere and spread, and consequently assume new shapes based on the balance of cohesive and adhesive forces. Morphogenetic processes like folding, invagination and evagination, and branching, not possible in purely liquid-like materials, occur when a basal lamina is present. In coordination with PCP, the basal lamina enables eumetazoans to form multilayered bodies, tubular structures, and appendages [60].

Although metazoan cell masses are novel kinds of materials with unique morphogenetic inherencies, these are only realized if more than one cell at a time undergoes polarization, expresses adhesive differences, or produces extracellular materials. In some cases, random assignment of cell states can lead to a non-random morphological outcome. This happens, for example, when cell adhesion occurs in salt-and-pepper pattern and cell sorting generates a phase-separated tissue (reviewed in [15]). More generally, local or global communication is required to coordinate cell behaviors to produce consistent outcomes. Coordination can occur chemically, by the action of *morphogens* (i.e., diffusible signal molecules), physiologically, by synchronization of biochemical oscillations, or bioelectrically, by tissue voltage gradients.

Wnt is a multi-functional morphogen produced by all animals, including placozoans and sponges. Like the Metazoa-defining cadherins, Wnt is a protein novelty, not found outside this group. By affecting tissue regions directly adjacent to the cells that secrete it, Wnt induces them to express certain genes, or to undergo ABP or PCP [61]. Other morphogens (some only found in certain subgroups of animals) affect gene expression in a concentration-dependent fashion [62]. The resulting cellular heterogeneity in early embryos and organ primordia is referred to as *pattern formation*. Elaborate variants are possible: as described by the mathematician Alan Turing [63], if two or more morphogens interact the patterns can be spatially periodic, as in the digits of the tetrapod limb, hair follicles, or pigment stripes (reviewed in [15]).

Morphogens spread by molecular diffusion, spanning a distance of ten cell diameters or less. Coordination of cell activities over 100 or more cell diameters occurs during some morphogenetic processes, however. This can be achieved if cells undergo periodic changes in a signal-reception factor. Such cellular oscillations can start out out-of-phase with one another but can spontaneously come into synchrony, making a broad region of tissue behave as a coordinated medium, a "morphogenetic field" [64-66]. With a suitable ratio of synchronous oscillations, morphogen diffusion, and growth, certain embryonic fields undergo a timed sequence of local boundary formation, organizing into *somites*, paired tissue blocks that are the precursors of the vertebrae [67]. The ability to sustain dynamical processes such as biochemical oscillations defines animal tissues as "excitable media" [68, 69], a property shared with some nonliving materials.

Bioelectical templating and restoration of tissue form are less understood mechanistically than either morphogen action or oscillation synchronization. Rather than determining the morphological motifs of developing animal tissues, it stabilizes, records, and (uncannily) resurrects them. This phenomenon of "anatomical homeostasis" is based on resting membrane potentials generated concomitantly with development [70]. It is mediated by transport of ions and other small molecules through gap junctions and tracks embryogenesis and organ formation in invertebrates and vertebrates, interactively mapping out the prospective locations of, for example, planarian heads, and amphibian faces and limbs [71]. Perturbation of these bioelectric fields by drugs, or manipulation of gap junction protein expression alters anatomy in predicable ways [72]. Most remarkably, establishing suitable resting potentials can guide the regeneration of lost body parts or induce ectopic structures [73].

The morphogenetic repertoire of metazoan cell masses marks them as a unique form of matter, with organizational propensities that are partly explicable by properties shared with nonliving counterparts. The metazoan tissue morphospace was further enhanced by the appearance of the basal lamina and other extracellular matrix (ECM) molecules. The gel-like and solid *connective tissues* that arose with the ECM, which are physically different from the liquid-like ones that coexist and interact with them during embryogenesis, enabled the formation of exo- or endoskeletons. The inherencies of the resulting composite tissues led the rise of more than 30 morphologically distinct animal body plans (reviewed in [57]).

Materials generate structures by physics, not by computation. Nonetheless, the conviction that there are genetic programs for embryogenesis has been a commonplace over the last half-century [6]. This idea is disconfirmed, however, by the fact that embryos of animals, including mammals, can be disassembled into separate cells part-way into development with each producing a normal individual [74]. A single mouse embryo constructed by fusing two with different sets of parents also develops normally [75]. The latter manipulation can even be performed with embryo cells of sheep and goats, members of different genera of a mammalian subfamily which are evolutionarily separated by approximately 4 million years [76]. It can also be done using embryos of zebrafish and medaka, members of different orders of teleost fish that diverged more than 300 million years ago and have very different developmental rates and timetables [77].

While a "normal" outcome is difficult to define in these biologically unprecedented interspecies chimeras, the morphologically intermediate individuals that result are viable and physiologically functional. To conclude that their embryos form according to computer-like genetic program one would have to imagine that two computers, running substantially different software, can be physically disassembled and combined, and implement on the spot a fully functional program yielding an amalgamated but serviceable output. The notion that embryos are formed by genetic programs is untenable.

**4. Mobilization of intrinsic cell functionalities in multicellular development**

Although there is no support for the idea that development of multicellular *form* is a computation-like process, the production of functionally *specialized cell types and organs* appears to incorporate certain elements of programming-type activity. As mentioned in Section 2, the cells of metazoan embryos use a mode of gene expression during development distinct from those of any other life-form [9]. In the remainder of this section I describe how this novel apparatus enables the embryo's cells to co-opt, accentuate, and partition gene co-expression networks that evolved in ancestral organisms [78], effectively making it an engine of amplification and partitioning of intrinsic cell functionalities [79].

The previously mentioned eukaryotic write-read-rewrite transcription regulatory system [44] is employed in the metazoan cell nucleus at biomolecular condensate-based foci or "hubs." These liquid-like domains contain scaffolding proteins and transcription factors, including those designated "pioneer" factors owing to their capacity to "open" chromatin domains, releasing genes from being otherwise "silenced" by structural and epigenetic factors [80]. Most notably, the hubs are sites of congregation of enhancer sequences, short stretches of DNA located often far from, but sometimes within, specific genes, and involved in amplifying their expression [81]. The expression mechanism also depends on non-protein coding RNA molecules transcribed from the enhancers: "eRNAs" [82].

In a scenario proposed by Arenas-Mena [83], the described metazoan-specific mode of gene regulation derives from the repurposing of the response of ancestral cells to environmental cues. He notes that non-metazoan eukaryotes regulate genes using distinct types of gene-adjacent promoter sequences. Some promoters control the expression of genes specifying "housekeeping" functions, ensuring levels of expression tied to maintenance of normal cell physiology. Others are "inducible", mediating cell response to external factors by selective upregulation of optionally expressed genes. Arenas-Mena suggests that the enhancer sequences of metazoan development-associated genes arose from inducible promoters of ancestral cells, which became linked to constitutive cell functionalities which could then be differentially expressed. This new regulatory architecture had the effect of bringing a modality that evolved to mediate interactions originating *outside* the (single-celled) individual to the *inside* of a new kind of (multicellular) individual. This enabled the construction of body plans containing specialized cell types.

An additional feature of this gene regulatory apparatus is DNA elements that tether genes separated over long chromosomal distances, enabling their coordinated expression [84]. "Super-enhancers" [85] are sites in which multiple enhancers mediate the elevated expression of genes organized into co-expression domains representing functional pathways [86]. The hierarchical logic of metazoan developmental gene regulation is reflected in "topologically associating domains" (TADs: [81]) which typically contain one or more genes with significant roles in development [87]. TAD-associated genes can initiate cell lineages by bringing together pioneer and lineage-determining transcription factors [88]. The latter initiate a series of state transitions in stem cells that culminate, via intermediate cell types, in terminally differentiated cells and specialized tissues.

Most of the genes recruited and amplified by these cell type-generating expression hubs can be traced to life-sustaining functionalities that were already present in the holozoan cell populations from which the metazoans arose: motility, electrical excitability, detoxification, lipid storage, oxygen capture and so forth. The appropriation of intrinsic single-cell functionalities for cell-type specialization is accompanied by suppression of other functionalities irrelevant to the differentiating cell (e.g., replication in skeletal muscle and neurons; motility in erythrocytes) (reviewed in [9]).

The metazoan gene regulatory apparatus incorporates the nucleic acid-based information storage and addressability common to all cells [43] and the histone-based write-read-rewrite capability that emerged with eukaryotes [44]. The chromatin-enhancer-based expression hub system, which enables "function-calling" [51] of multigene pathways in metazoan cells, is a major factor in the emergence of increased phenotypic complexity during animal evolution. Among the animal groups, the vertebrates exhibit a substantial increase relative to the invertebrates in the number and connectivity of developmentally regulated genes. This is associated with higher numbers of regulatory sequences addressable by developmental cues and a significant rise in open chromatin domains with increased density of enhancers [89].

Consistent with this, the early-diverging, enhancer-lacking placozoans have about 9 functionally simple cell types [90], and sponges about 18 [91], arthropods about 250 [92], and humans more than 400 [93] functionally complex types. Thus, while the morphospace of metazoan anatomy (see Section 3) was largely explored and exhausted millions of years ago, addition of cell types based on new appropriations of latent functionalities continued well past that time.

Up to this point we have considered inherencies of unitary forms of matter (generic, biogeneric, and physically exotic) and internally heterogenous non-unitary organized entities (e.g., cell themselves) as determinants of biological properties. In Section 3, I presented embryological arguments that morphogenesis, and development in general, are not generated computationally. From the evidence presented in this section, the metazoan gene regulatory apparatus is also an unsuitable medium to conduct computations, due to the volatility of strength and topology of network links owing to the intrinsic disorder of developmental transcription factors and other components of the expression hubs [94], and the variability in the deployment of enhancers [95]. This contradicts the popular claim over the past half-century that cell differentiation is the set of outputs of "genomic computers" constructed of switchboard-like arrays of gene regulatory networks (GRNs) [96, 97].

A different model for cell differentiation, also advanced before the discovery of the novel nature of metazoan developmental gene regulation, was based on formal properties of dynamical systems (see [9] for a review). Dynamical systems are not themselves forms of matter, but mathematical representations of idealized networks of mutually influencing components. Typical formats are coupled ordinary differential equations (ODEs), with continuous-valued variables, or networks of logical (e.g., Boolean) functions with discrete-valued variables. If a material system exhibits conserved topology of causal connections, stoichiometry (i.e., changes among reactants occurring in fixed proportions), and conservation of mass (reaction rates are proportional to concentrations of components, it can be modeled by an ODE or Boolean network with deterministic outcomes. Systems of coupled chemical reactions are a classic form of matter that meet the criteria for dynamical systems modeling [98].

The dynamical systems model of cell differentiation was originally based on simulations of randomly constructed Boolean networks and contained the implication that generation of multiple cell types was a mathematical inevitability [99]. The main idea was that networks of regulatory genes that mutually control one another's transcriptional activity can be adequately modeled by systems of Boolean functions or ODEs. Such systems typically do not have one stationary state (defined by the joint values of each network component when all rates of change are in balance) but exhibit a multiplicity of point or oscillatory "attractors." The system will come to rest at one or another of these attractor states depending on initial conditions, or jump between them with suitable perturbations (e.g., developmental signals).

Through a series of mathematical refinements [100, 101] the original concept that "it is almost an inevitable hypothesis that the distinct cell types of an organism correspond to the distinct attractors of the network" [8] continues to prevail. A more recent statement asserts that "one genome specifies one particular GRN wiring diagram, which in turn specifies one particular landscape that captures that genome's entire developmental potential" and that "the genome determines developmental trajectories and terminal gene expression profiles of cell types and thus, ultimately, the organismal phenotype" ([102], p. 155).

Such multi-attractor models have proved useful for describing transitions between adjacent cell states in developmental lineages, where states are functionally related and the switch between them is typically controlled by one or a few parameters [103, 104]. However, they cannot plausibly account for the full set of differentiated cell types (ranging, as noted above, from less than 10 to several hundred) of any animal species. Briefly, the number of attractors in a discrete or continuous dynamical system, and the values of the variables (regulatory factors in these cases) at each of the attractor states, are purely mathematical effects of the system connectivity and rate parameters. There is a vanishingly small probability that all the system states of such networks, whether randomly built or specifically evolved, represent a mutually compatible set of cell functions constituting physiologically viable tissues (contractile muscles, supportive skeletons, excitable nerves and so forth) and organs (e.g., hepatocytes and bile ducts in liver, oxygen-permeable and surfactant-forming cells in lung). Even if this highly unlikely situation had pertained at an early stage of evolution, once the number of regulatory genes had increased with phylogenetic diversification it would have been mathematical [105, 106] and biological near impossibilities that the larger networks would have retained the original attractor set (necessary, since basal functions are conserved) while adding new attractors that embodied mutually compatible and well-integrated functions that did not previously exist; see [9] for further discussion).

The multi-attractor dynamical system models for cell differentiation thus misidentifies the generic properties of mathematical objects with supposed inherent properties of biological systems. In particular, the embryo cell chromatin of an animal species and its organization into phase-separated enhancer-dependent chromatin hubs is not a unitary form of matter with definable intensive properties or other inherencies. Instead, the generation of species-characteristic cell types is a heterogenous process that draws on a stored set of ready-made cellular functionalities in the form of evolutionarily deep gene coexpression networks. It deploys them in a fashion that (owing to the information storage and processing capabilities of the constituent cells) is quasi-computational, but unlike a dynamical system.

The evolution of animals proceeded within constraints due to limited possibilities of forms and functions (respectively, inherent, and intrinsic). As these were played out, the arena changed to subsystems such as the nervous system, and the intensification and elaboration of their functions. The diversity of neuronal cell subtypes in the mouse brain is on the order of 5000, for example [107]. Is brain tissue a connectionist computer, a unitary form of matter with mentality-supporting inherencies and system dynamics, composites of these, or none of the above? The next section will explore this question.

## 5. Material inherencies, computation, and mind

Although there are ongoing debates about what kind of computer the mind (or brain) is – digital, connectionist [108], Universal Turing [109] – there is a widespread commitment (with notable exceptions [110]), among neuroscientists and cognitive philosophers to a computationalist notion of mental activity. A recent book cites the philosopher Daniel Dennett, for example, as "believing that computations are the functional equivalents of thoughts" [111]. The view, advanced in the 1970s, that brains instantiated the capabilities of digital computers, has been broadly rejected (reviewed in [112]), but the the idea of the brain as some kind of computer still reigns. Piccinini, who advances connectionism and machine learning ("artificial intelligence") as computationalist alternatives, notes, for example, that

> [w]ithin conventional computers, the only kind of information processing that takes place is the computation of outputs based on inputs and internal states. In contrast, neurocomputational systems are constantly engaged in two types of information processing at once. Like conventional computers, they yield outputs as a function of their inputs and internal states. Unlike conventional computers, they also learn—that is, they use a number of information sources together with their self-organizing capacity to alter their structure and, therefore, their future functions ([113], p. 8).

Piccinini also makes "neural structural representations" central to his proposal for how the computational brain creates semantic content, i.e., meanings that can be manipulated as information. He defines representations of this sort in computer parlance, as "a state of a simulation of a target, where a simulation is a system of states, homomorphic to their target, which can evolve to match the evolution of their target to some degree of approximation" ([113], p 5).

Representationalism, the idea that cognition is the manipulation of mental representations is marshalled against "enactivism," an anti-computationalist perspective that denies that representations of this, or any other sort, figure into mental activity. Part of this stems from the fact that the concept of autopoiesis (the inspiration for the enactivist approach; see Section 2) "replaced the idea of mental representations with the idea that the environment triggers perturbations to the ongoing and operationally closed processes of the living system" ([110], p. 19).

More important to the rejection by some cognitive philosophers of both computationalism and representationism (as overarching theories of the mind) are difficulties in pinning down what either concept is intended to explain. Milkowski [108] notes ambiguities in the computational modes (digital, analog, hybrid, connectionist) posited to underlie specific mental processes, and the difficulties of identifying the physical systems that implement the respective computations. Questions he raises include: Can the distributed representations afforded by connectionist models do the work of the node-based representations of (less biologically plausible) digital models? Do the artificial neural networks of connectionist models of cognition correspond to networks of living neurons?

Concerning representationism, Favela and Machery [114] conducted a survey of more than 700 psychologists, neuroscientists, and philosophers and found that "researchers exhibit confusion about what counts as a representation and are uncertain about what sorts of brain activity involve representations or not." In perceptual systems, however, where there is some consensus on the existence of representations, their relation to the "wiring diagram" of the brain, which would seem to be necessary for their basis in neural computation, is challenged by the recent finding of "representational drift" in the olfactory cortex [115]. This is a phenomenon in which "the distribution of neural populations activated in response to an odorant (odorous chemicals) can shift entirely in the course of a month, suggesting that smells cannot be modeled via stable neural correlates" [116]. Similar effects occur in the visual cortex when stimulated with complex naturalistic stimuli [117], and most strikingly with the storage of memories, where the constant element in the drifting localization of the representation across brain areas is the neuron-generated electric field pattern [118], an echo of the anatomical homeostasis phenomenon [70] described in Section 3.

Another notion with currency in cognition studies is dynamical systems theory, which we encountered in models of cell differentiation (see Section 4). It is a central feature of the decades-old cognitive philosophy perspective known as the "dynamical hypothesis" [119, 120], but the term is used in a looser fashion in the science and philosophy of cognition than in the models advanced to explain cell differentiation. Beer [121] provides a general mathematical definition of dynamical systems, but due to the heterogeneity of the processes involved, he is unable to identify a signature kind of dynamics to cognition and concludes that "[o]n its own, the mere mathematical idea of a dynamical system is too weak to serve as a scientific theory of anything, and dynamical approaches within cognitive science are too rich and varied to be subsumed under a single 'dynamical hypothesis'." Put differently, if there was a prospect that the brain of any animal, or a functionally relevant portion of it, had a unitary material nature that could be modeled by a dynamical system, it has not been realized in any research program to date.

It seems possible that cognitive computationism is an example of the "Dürer's rhinoceros" phenomenon mentioned above, that is, misapplication of a known concept or technology to something puzzling. Representationalism may point to something real but perhaps more diverse in its material bases than a simple computationally based definition can capture. Dynamical systems of one kind or another may eventually explain aspects of mental function, but only after better characterizations of the material nature of the tissues in which they occur are available.

As an outsider to cognitive science, I will not offer solutions to any of these questions. In the brief comments that follow, however, I attempt to draw some lessons from the growth in understanding of embryonic development over the past half-century and the dispelling of some conceptual confusions of the past. To recall, the factors and terms of reference that proved helpful in these areas were:

(i) *Internal organizational properties of animal cells* that enable their life-sustaining functionalities (motility, excitability, nutrient uptake, waste excretion, detoxification, etc.). Included in these functionalities are (a) the transgenerational nucleic acid-protein code-like correspondences shared with all cellular life, (b) the eukaryotic-specific, within-generation, histone-based, write-read-rewrite record-keeping capability, and (c) the metazoan-specific cell-functionality appropriation and suppression apparatus used in cell differentiation.

(ii) *Generic and biogeneric material properties of multicellular clusters*. The inherencies of these unitary materials (e.g., morphological motifs of liquid-like and liquid + solid tissues) are understood from known physical principles, and their contributions to the evolution and development of body plans and organ forms (e.g., tissue layering, segmentation) are thereby comprehensible. Even if the originating morphogenetic processes are no longer developmentally efficacious, organismal anatomies still bear their stamp.

(iii) *Non-generic unitary materials*. These can be subcellular (e.g., biomolecular condensates) or multicellular (e.g., developing plant tissues). While they do not have counterparts in the nonliving world, they can be deconstructed and analyzed physically, and their inherencies largely (plant tissues) or eventually (biomolecular condensates) understood.

(iv) *Scaffolding effects in unitary biomaterials*. These can be either generic (e.g., synchronous oscillations of regulatory gene expression in developing tissues) or nongeneric, i.e., found only in living systems (e.g., bioelectric restoration of morphology).

(v) *Hybrid and composite assemblages acting at the multicellular scale*. These bring together one or more of the factors in (i)-(iv) to perform a reproducible organismal role. The differentiation of specialized cell types by animal embryos is an example.

Items (i) and (v) (but not (ii-iv)) embody quasi-computational capacities, i.e., storage and manipulation of protein synthesis-relevant information, storage, and manipulation of "experiences" of the biochemical environment, storage, and manipulation of cell functionalities. None of these activities make their (cellular or multicellular) vehicles computers or computer-like.

Concerning the brain, the first important thing to note is that the fundamental things it does are already present in cells (implicitly included among the functionalities under item (i), above). Lyon defines cognition as

> …the sensory and other information-processing mechanisms an organism has for becoming familiar with, valuing, and interacting productively with features of its environment [exploring, exploiting, evading] in order to meet existential needs, the most basic of which are survival/persistence, growth/ thriving, and reproduction ([122], p. 416).

She proceeds to cite a vast body of literature that locates all these processes in bacteria, single-celled eukaryotes, plants, fungi, and other organisms without nervous systems (see also [45]). Whether it is termed, "proto-," "minimal," or "basal" cognition, once cells had appeared, many of the capabilities that later came to characterize nervous systems and brains were already in place. Further, lacking networks of neurons, single-celled organisms and communicating populations of them, e.g., biofilms [123]), cannot conceivably perform these activities by acting as computers, connectionist or otherwise.

The earliest emerging animals – placozoans and sponges – also lack neurons, but nonetheless also "interact productively" with their environments. Their capacity to do so goes beyond those of their single-cell ancestors, owing to the fact that their more complex bodies provide them with novel structural and functional enablements [49] and corresponding ecological affordances, in Gibson's sense of environmental features with organism-specific meanings [124]).

The first eumetazoans, the ctenophores and cnidarians, with simple two-layered bodies, had (and their present-day descendants have) nerve nets that coordinate whole-body contractile activity, feeding and sensation. Some of these behaviors are retained in hydra that lack nerves, which can also live for years if manually fed. There is no indication that the functions mediated by the nerve net in normal hydra are accomplished computationally [125].

Planaria, three-layered flatworms, have a rudimentary brain and two parallel nerve cords. They can execute complex behaviors, remember experiences, and exhibit cohort-specific "habits" (e.g., nocturnality). Some of these patterns of activity are expressed in portions of the animal's body lacking the brain, almost immediately after they are isolated and days before regeneration (a capacity planaria are famous for) has occurred [126, 127]. Once again, while brains facilitate life-sustaining behaviors, they are not obligatory and thus do not perform computations or operate as dynamical systems in the conventional senses.

This suggests that brains and other nervous systems may be "add-ons" to somatic-based functions (sensory, behavior coordinating, and cognitive) that derive from ancestral cellular properties (see also [128]). In this view, while brains may be essential for higher cognitive activities like sentience and planning [111]), their contributions to these functions could be as forms of matter, or composites of such materials (items (iii)-(v), above), with various inherent modes of activity, serving the bodies with which they coevolved analogously to other organs like the pancreas or skeletal muscles, not as the central processing unit of the self. Since animals were cognitive before they had brains, the neuronal circuits or primitive brains early in evolution were likely optional enhancements. In the lineages in which they appeared, after millions of years they are no longer so, but have become entrenched, functionally and generatively [129].

The brain-as-novel-form-of-matter concept was the inspiration for Karl Pribram's proposal of the "holographic brain" [130]. There are several versions of the theory which variously invoke quantum effects or biogeneric emulations of such effects, and it warrants updating in light of newer knowledge of brain tissue properties and dynamics. The focus of such models on inherent and intensive properties of brain matter (rather than wiring diagrams), however, suggest bases for puzzling observations of retention of learned activity patterns and other memories when brains undergo remodeling during regeneration or metamorphosis [131], or when large portions of even the human brain are missing [132].

Just as the developing embryo employs the gene-addressing, write-read-rewrite experience-recording, and functionality-calling capabilities intrinsic to its constituent cells (see Sections 2 and 4, above), the exotic forms of matter constituting brains may exhibit some quasi-computational activities. Symbol manipulation, a generally accepted hallmark of mental activity, could depend on inherencies of brain-matter, such as the recording of visual [133] and auditory [134] experiences. These are representations, but not in the computational sense. That they can be stored and replayed at later times, voluntarily or involuntarily (e.g, the poet Wordsworth's "inward eye" [135]; "earworms" [136]) suggests that the recording media might be analog, like photographic film or magnetic tape, which are more plausible as an inherent tissue property than either digital storage media or persistent activity patterns of a neural network.

Like the embryonic tissues that form somites or digits (see Section 2), brain matter may perform some of its functions (e.g., consolidation of memories) by virtue of being an excitable medium capable of sustaining synchronous oscillations [137, 138]. The idea is that biological matter can be a venue of dynamical processes without (as postulated by the dynamical hypothesis for cognition [120]) being itself a dynamical system.

Of more general relevance, however, is the question posed by Baluška and Levin, "How does biological matter give rise to decision-making, memory, representation, and goal-directed activity?" [128]. Given the known steps in animal evolution, the way brains function is only part of the answer. The approach sketched out here is motivated by the conclusion (described above and earlier, e.g., [16, 49]) that the material properties of biological systems are decisive for their specific developmental and physiological properties and evolutionary trajectories. It therefore rejects what is referred to as "multiple realizability" of cognitive systems (the doctrine that "a single mental kind (property, state, event) can be realized by many distinct physical kinds" [139]) in favor of an intensification, down to the histological level, of the perspective known as "embodied cognition" [140].

## 6. Conclusion

In this paper I have appealed to specific properties of cells and complex biological materials at a variety of scales to mount a critique of metaphorical computationalism and dynamical systems approaches to understanding multicellular development and its evolution. I have primarily focused on morphogenesis and cell differentiation, and the generation of patterned arrangements of cells in the metazoan animals. I have also drawn lessons from the history of these fields that might be helpful to cognitive scientists in avoiding pitfalls as they approach their own complex subject.

Much of the information presented here was not available until recently, leading theoreticians to sometimes dress poorly understood phenomena in the armor of known concepts, what I refer to as the "Dürer's rhinoceros" effect. A persistent theme in the past has been attribution of computer-type algorithms ("genetic programs of development") and hardware ("the genomic computer"), and properties of complex mathematical systems ("dynamical landscapes," "attractors"), to processes that have turned out to be other things entirely (inherencies of unitary materials, mobilization of intrinsic functionalities of single cells). This does not imply that computational and dynamical system modalities have no role in theoretical biology. The descriptions above indicate that organ primordia of developing embryos put in "calls" of stored single-cell functionalities (or duplicates of their gene co-expression networks) in constructing specialized cells. Embryonic tissues can also serve as excitable media capable of sustaining dynamical processes such as Turing-type pattern formation and collective oscillatory phenomena.

The cells that constitute these tissues also perform quasi-computational activities: addressing stored information for constructing proteins, recording environmental encounters via revisable histone marks affecting subsequent gene expression. Metazoan cells can also mobilize arrays of enhancers to quantitatively modulate expression of key genes and coexpression pathways, an activity more gadget-like (though employing functionally enigmatic biomolecular condensates) than dynamical.

The animal embryo is thus a congeries of biogeneric materials composed of functionally complex subunits containing stored ancestral and experiential information. To attain its mature form, the embryo selectively recruits cell functionalities evolutionarily based on ancestral ones. Given this, it is difficult to conceive of development, or even a program of developmental instructions, as multiply realizable, that is, encoded formally in other media than those described.

By extension, understanding the cognitive capacities of animals must begin with their constituent cells, which as discussed above (Section 5), already had them in a basal form before the evolution of brains or even neurons. The "reafferent" cognition of aneural animals such as placozoans, in which an organism responds to the consequences of it own actions [141], can only be interpreted as collective properties of their cells. These properties are fortified by enablements such as morphological embellishments based on material inherencies, and sensory and gustatory organs developmentally derived from the cells' store of intrinsic functionalities. In placozoans, these amount to relatively simple features like concerted motions of cilia and primitively specialized cells (since placozoans lack enhancers) such as gravity sensors (reviewed in [49]).

The structural and functional enablements of more complex animals, including their brains, make more sophisticated kinds of cognition possible [111]. But it is significant that even simpler multicellular systems, like the artificially constructed "biobots" of Levin, Bongard and coworkers [142], in which spheres of aggregated cells derived from frog embryos perform life-sustaining activities unrelated to those of their originating species or any other known animals, in the absence of morphological or cell specializations. This highlights the role (and cellular foundation) of the self-motivated activities known collectively as *agency*, without which all the morphological and functional novelties generated by the inherencies described above, and the organisms that bear them, would be dead in the water [143]. Since everything about living cells and the organisms comprised of them suggest that they are hedonic, adventurous agents, it is difficult to credit the idea that the main function of animal cognition is (as has been proposed) something as pedestrian as performing computations that "minimize surprise" [144].

I conclude with a speculative extrapolation of the developmental processes described above to an enigmatic aspect of brain function, localist (single cell) representations of complex features and category concepts, aka "grandmother cells." There is presumptive evidence for such representations [145], although the concept has been disputed, and the phenomena attributed instead to multi-neuron "sparse coding" circuits [146]. If they exist, concept cells could account for the persistence of memories through metamorphosis [131] and paradoxical disparities between a brain's tissue volume and its assumed information content [132] discussed in Section 5. The main defense of the notion has been based on cognitive computationalism, e.g., "[T]he brain is a massively parallel, distributed computing system that is symbolic" [145]. Here, for argument's sake I will assume that such cells, or something like them, exist, and present a (non-computationalist, living matter-based) hypothesis for their genesis.

In Section 4, I discussed Prohaska and coworkers' characterization of the write-read-rewrite system of histone modifications that record and revise within-lifetime chemical experiences of eukaryotic cells [44]. I also described Arenas-Mena's proposal on the recruitment of ancestral environment-responsive promoters in the evolution of enhancers used in gene (later recognized to be co-expression network) expression amplification during metazoan development [83]. My suggestion is that a cell responsive to complex stimuli could be produced when a version (perhaps brain-specific) on the Arenas Mena mechanism gathers the activity patterns of a sparse-coding network (possibly employing biomolecular condensates at the dendritic spines of the respective neurons [40, 41]) and funnels them into an individual cell. The target cell, employing its revisable recording capabilities, would thus become subfunctionalized into one embodying the full circuit's memory.

This hypothesis is experimentally testable. Even if it is falsified, it is just one of many possible conjectures consistent with known properties of the brain's tissue and cells, that is, its material nature, and not its metaphorical parallels to general purpose information-processing machines.

**Acknowledgment**

The author gratefully acknowledges the financial support of the John Templeton Foundation, United States (#62220). The opinions expressed in this paper are those of the authors and not those of the John Templeton Foundation.